\documentclass[3p,times,procedia]{elsarticle}
\usepackage{nupha_ecrc}
\usepackage{color}
\usepackage{graphicx}
\usepackage{dcolumn}
\usepackage{bm}
\usepackage{subfigure}
\usepackage{slashed}
\usepackage{marvosym}
\usepackage{textcomp}
\usepackage{threeparttable}
\usepackage{booktabs}
\usepackage{setspace}
\usepackage{latexsym}
\usepackage{mciteplus}
\usepackage{amsmath}
\usepackage[margin=20pt,small]{caption}

\volume{00}

\firstpage{1}

\journalname{Nuclear Physics A}

\runauth{}

\jid{nupha}

\jnltitlelogo{Nuclear Physics A}

\usepackage{amssymb}

\usepackage[figuresright]{rotating}

\begin{document}

\begin{frontmatter}



\dochead{XXVIIth International Conference on Ultrarelativistic Nucleus-Nucleus Collisions\\ (Quark Matter 2018)}

\title{Non-Equilibrium Quantum Transport of Chiral Fluids from Kinetic Theory}


\author{Yoshimasa Hidaka$^{1,2}$, Shi Pu$^{3}$, Di-Lun Yang$^{1}$}

\address{$^1$Quantum Hadron Physics Laboratory, RIKEN Nishina Center,
	RIKEN, Wako, Saitama 351-0198, Japan.\\
	$^2$iTHEMS Program, RIKEN, Wako, Saitama 351-0198, Japan.\\
	$^3$	Department of Physics, The University of Tokyo,
	7-3-1 Hongo, Bunkyo-ku, Tokyo 113-0033, Japan.}

\begin{abstract}
We introduce the quantum-field-theory (QFT) derivation of chiral kinetic theory (CKT) from the Wigner-function approach, which manifests side jumps and non-scalar distribution functions associated with Lorentz covariance and incorporates both background fields and collisions. The formalism is utilized to investigate second-order responses of chiral fluids near local equilibrium. Such non-equilibrium anomalous transport is dissipative and affected by interactions. Contributions from both quantum corrections in anomalous hydrodynamic equations (EOM) of motion and those from the CKT and Wigner functions (WF) are considered in a relaxation-time approximation (RTA). Anomalous charged Hall currents engendered by background electric fields and temperature/chemical-potential gradients are obtained. Furthermore, chiral magnetic/vortical effects (CME/CVE) receive viscous corrections as non-equilibrium modifications stemming from the interplay between side jumps, magnetic-moment coupling, and chiral anomaly. 
\end{abstract}

\begin{keyword}
Chiral Kinetic Theory, Chiral Anomalies, Weyl Fermions, Chiral Fluids	


\end{keyword}

\end{frontmatter}


\section{Introduction}
\label{}
In recent years, there have been increasing studies upon CKT as a relativistic quantum kinetic theory incorporating the chiral anomaly, which delineates the dynamics of weakly coupled Weyl fermions both in and our of thermal equilibrium. Despite the validity at strongly coupled quark gluon plasmas (QGP), CKT could be applied to study the anomalous transport such as CME/CVE in heavy ion collisions (HIC) \cite{Vilenkin:1979ui, Kharzeev:2007jp,Fukushima:2008xe,Landsteiner:2011cp}. The construction of CKT was initiated in Refs.\cite{Son:2012wh, Stephanov:2012ki} via a semi-classical approach. In the adiabatic limit, the equations of motion and phase space measure of Weyl fermions are modified by the quantum correction from Berry curvature, which results in a kinetic theory involving the chiral anomaly. Moreover, it has been found later that the distribution function in the CKT follows modified frame (Lorentz) transformation associated with side-jump phenomena \cite{Chen:2014cla,Chen:2015gta}. Such a side-jump effect plays an important role to maintain Lorentz covariance of the CKT and generate the CVE. Nevertheless, the aforementioned studies only consider the case in the absence of background electric/magnetic fields. 
          
On the other hand, it is also desired to derive a covariant CKT from QFT.
Although there have been previous attempts on such first-principle derivations in Refs\cite{Chen:2012ca,Son:2012zy}, the studies are limited to the conditions in a steady state or at large chemical potentials and collisions are not included. Recently, a comprehensive derivation in QFT for the covariant CKT involving both background fields and collisions has been achieved in Refs.\cite{Hidaka:2016yjf,Hidaka:2017auj} via the Wigner-function approach. In addition, the formalism is further implemented to analyze second-order quantum transport of chiral fluids \cite{Hidaka:2017auj,Hidaka:2018ekt}, where novel anomalous Hall currents and viscous corrections upon CME/CVE are found. In this proceeding, we will review the works in Refs.\cite{Hidaka:2016yjf,Hidaka:2017auj,Hidaka:2018ekt} for recent progress in CKT and the applications in non-equilibrium anomalous transport.

\section{Wigner Functions and Chiral Kinetic Theory}    
WF as the phase-space distributions of lesser(or greater) propagators are defined in terms of the Wigner transformation,
\begin{eqnarray}\label{WF_def}
\grave{S}^{<}(q,X)\equiv\int d^4Ye^{\frac{iq\cdot Y}{\hbar}}\langle\psi^{\dagger}(y)\psi(x)\rangle,
\end{eqnarray}
where $\psi$ denotes the fermionic-field operator and $Y=x-y$ and $X=(x+y)/2$. Also, the gauge link is implicitly embedded and $q_{\mu}$ thus represents the kinetic momentum. We will focus on just right-handed fermions hereafter. The dynamics of WF is dictated by Kaddanof-Baym(KB)-like equations obtained from the Dyson-Schwinger equations.
By parameterizing $\grave{S}^<=\bar{\sigma}^{\mu}\grave{S}^<_{\mu}$, up to $\mathcal{O}(\hbar)$, the perturbative solution takes the form \cite{Hidaka:2016yjf},
\begin{eqnarray}\label{Wigner_S}
\grave{S}^{<\mu}(q,X)
=2\pi\bar{\epsilon}(q\cdot n)\Big(q^{\mu}\delta(q^2)
+\hbar\delta(q^2)S_{(n)}^{\mu\nu}\mathcal{D}_{\nu}
+\hbar\epsilon^{\mu\nu\alpha\beta}q_{\nu}F_{\alpha\beta}\frac{\partial\delta(q^2)}{2\partial q^2}
\Big)f^{(n)}_q,
\end{eqnarray}
where $\bar{\epsilon}(q\cdot n)$ represents the sign of $q\cdot n$ and
$S^{\mu\nu}_{(n)}=\epsilon^{\mu\nu\alpha\beta}q_{\alpha}n_{\beta}/(2q\cdot n)$
corresponds to the spin tensor depending on a frame vector $n^{\mu}$ stemming from the choice of a spin basis. Here we denote $\mathcal{D}_{\beta}f^{(n)}_q=\Delta_{\beta}f^{(n)}_q-\mathcal{C}_{\beta}$, where $\Delta_{\mu}=\partial_{\mu}+F_{\nu\mu}\partial^{\nu}_q$,  $\mathcal{C}_{\beta}=\Sigma_{\beta}^<\bar{f}^{(n)}_q-\Sigma_{\beta}^>f^{(n)}_q$ with $\Sigma_{\beta}^{<(>)}$ being lesser/greater self-energies and $f^{(n)}_q$ and $\bar{f}^{(n)}_q=1-f^{(n)}_q$ being the distribution functions of incoming and outgoing particles, respectively. Note that $f^{(n)}_q$ can be either in and out of equilibrium. The $\mathcal{O}(\hbar)$ terms in (\ref{Wigner_S}) give rise to quantum corrections, in which the last term modifying the dispersion relation yields the CME in equilibrium and the $S^{\mu\nu}_{(n)}$-dependent term leads to side jumps and the modified frame (Lorentz) transformation. Since WF constructed from (\ref{WF_def}) is frame independent, the side-jump term accordingly infers the frame transformation for the distribution function \cite{Hidaka:2016yjf}, $f^{(n')}_q=f^{(n)}_q+\hbar(q\cdot n')^{-1} S^{\mu\nu}_{(n)}n'_{\nu}\mathcal{D}_{\nu}f^{(n)}_q$.
Such frame transformation preserves the frame independence (Lorentz covariance) of $\grave{S}^{<\mu}$ up to $\mathcal{O}(\hbar)$.

From the KB-like equations and WF in (\ref{Wigner_S}), we further derive the CKT up to $\mathcal{O}(\hbar)$ \cite{Hidaka:2016yjf,Hidaka:2017auj},
\begin{eqnarray}\label{CKT}
\delta\Bigl(q^{2}-\hbar \frac{B\cdot q}{q\cdot n}\Bigr)\big(\Box(q,X)f^{(n)}_q-\mathcal{C}_{\text{full}}\big)=0,
\end{eqnarray}
where 
\begin{eqnarray}
\Box(q,X)=\Big[
q\cdot\Delta+\hbar(q\cdot n)^{-1}S_{(n)}^{\mu\nu}E_{\mu}\Delta_{\nu}
+\hbar S_{(n)}^{\mu\nu}(\partial_{\mu}F_{\rho \nu})\partial^{\rho}_{q}
+\hbar(\partial_{\mu}S^{\mu\nu}_{(n)})\Delta_{\nu}\Big],
\end{eqnarray}
and 
\begin{eqnarray}\label{C_full}
\mathcal{C}_{\text{full}}=\Bigg(q^{\mu}+\hbar (q\cdot n)^{-1}S_{(n)}^{\nu\mu}E_{\nu}+\hbar\big(\partial_{\rho}S^{\rho\mu}_{(n)}\big)\Bigg)\tilde{\mathcal{C}}_{\mu},\quad 
\tilde{\mathcal{C}}^{\mu}=\mathcal{C}^{\mu}+\hbar\frac{\epsilon^{\mu\nu\alpha\beta}n_{\nu}}{2q\cdot n} 
\big(\bar{f}^{(n)}_q\Delta^>_{\alpha}\Sigma^{<}_{\beta}-f^{(n)}_q\Delta^<_{\alpha}\Sigma^{>}_{\beta}\big)
.
\end{eqnarray}
Here, we define $B^{\mu}$ and $E^{\mu}$ by decomposing the field strength into $F_{\alpha\beta}=-\epsilon_{\mu\nu\alpha\beta}B^{\mu}n^{\nu}+n_{\beta}E_{\alpha}-n_{\alpha}E_{\beta}$. The CKT now also incorporates the quantum corrections in collisions. By solving $f^{(n)}_q$ from (\ref{CKT}) and obtain the WF in (\ref{Wigner_S}), one can unambiguously compute physical quantities defined in QFT such as the energy-momentum tensor and charge currents, $T^{\mu\nu}=\int_q\big(q^{\mu}\grave{S}^{<\nu}+q^{\nu}\grave{S}^{<\mu}\big)$ and $J^{\mu}=2\int_q \grave{S}^{<\mu}$, where $\int_q=(2\pi)^{-4}\int d^4q$.

\section{Second-Order Anomalous Transport of Chiral Fluids} 
The formalism introduced in the previous section is then implemented to study non-equilibrium quantum transport of chiral fluids near local equilibrium based on the gradient expansion and RTA \cite{Hidaka:2017auj,Hidaka:2018ekt}. Technically, the strategy is to solve for the non-equilibrium distribution function with the small deviation from the equilibrium one, $\delta f_q=f^{(u)}_q-f^\text{eq}_q$. It is found in Ref.\cite{Hidaka:2017auj} that the local equilibrium distribution function up to $\mathcal{O}(\hbar)$ in the co-moving frame $n^{\mu}=u^{\mu}$, where $u^{\mu}$ denotes the fluid velocity, takes the form $f^\text{eq}_q=(e^{g}+1)^{-1}$ with
$g=(q\cdot u-\mu+\hbar(\omega\cdot q)(2q\cdot u)^{-1})/T$ for $T$ and $\mu$ being the local temperature and chemical potential, respectively. Also, the vorticity $\omega^{\mu}$ is defined as $\omega^{\mu}\equiv \epsilon^{\mu\nu\alpha\beta}u_{\nu}\big(\partial_{\alpha}u_{\beta}\big)/2$. As shown in \cite{Hidaka:2017auj}, the $f^\text{eq}_q$ and corresponding WF in local equilibrium yield CME and CVE in charged currents and the associated non-dissipative transport in the energy=momentum tensor. The results therein agree with previous findings from distinct approaches such as anomalous hydrodynamics with thermodynamic constraints \cite{Son:2009tf,Neiman:2010zi}. Because CME and CVE are both at $\mathcal{O}(\hbar\partial)$, it is expected that the non-equilibrium quantum corrections appear at $\mathcal{O}(\hbar\partial^2)$.

In order to acquire an analytic result for non-equilibrium transport capturing qualitative features without the loss of generality, the RTA is applied by approximating 
$\mathcal{C}_\text{full}\simeq-(q\cdot u\delta f_q)/\tau_R$, where the constant $\tau_R$ denotes a relaxation time charactering the inverse strength of interactions. More detains of the Lorentz invariance and frame transformation of the RTA are discussed in Ref.\cite{Hidaka:2018ekt}. In addition to solving $\delta f_q$ from the CKT, we have to further introduce the hydrodynamic EOM led by the charge and energy-momentum conservation (with the chiral anomaly), $\partial_{\mu}J^{\mu}=-\hbar E_{\mu}B^{\mu}/(4\pi^2)$ and $\partial_{\mu}T^{\mu\nu}=F^{\nu\mu}J_{\mu}$,
which constrain the temporal derivatives of thermodynamic parameters. However, according to the matching conditions given by the CKT with the RTA, the non-equilibrium charge density and energy-density current should vanish. We are consequently able to define the equilibrium temperature and chemical potentials in a consistent manner. 

\begin{figure}[t]
	\begin{center}
		{\includegraphics[width=8cm,height=4cm,clip]{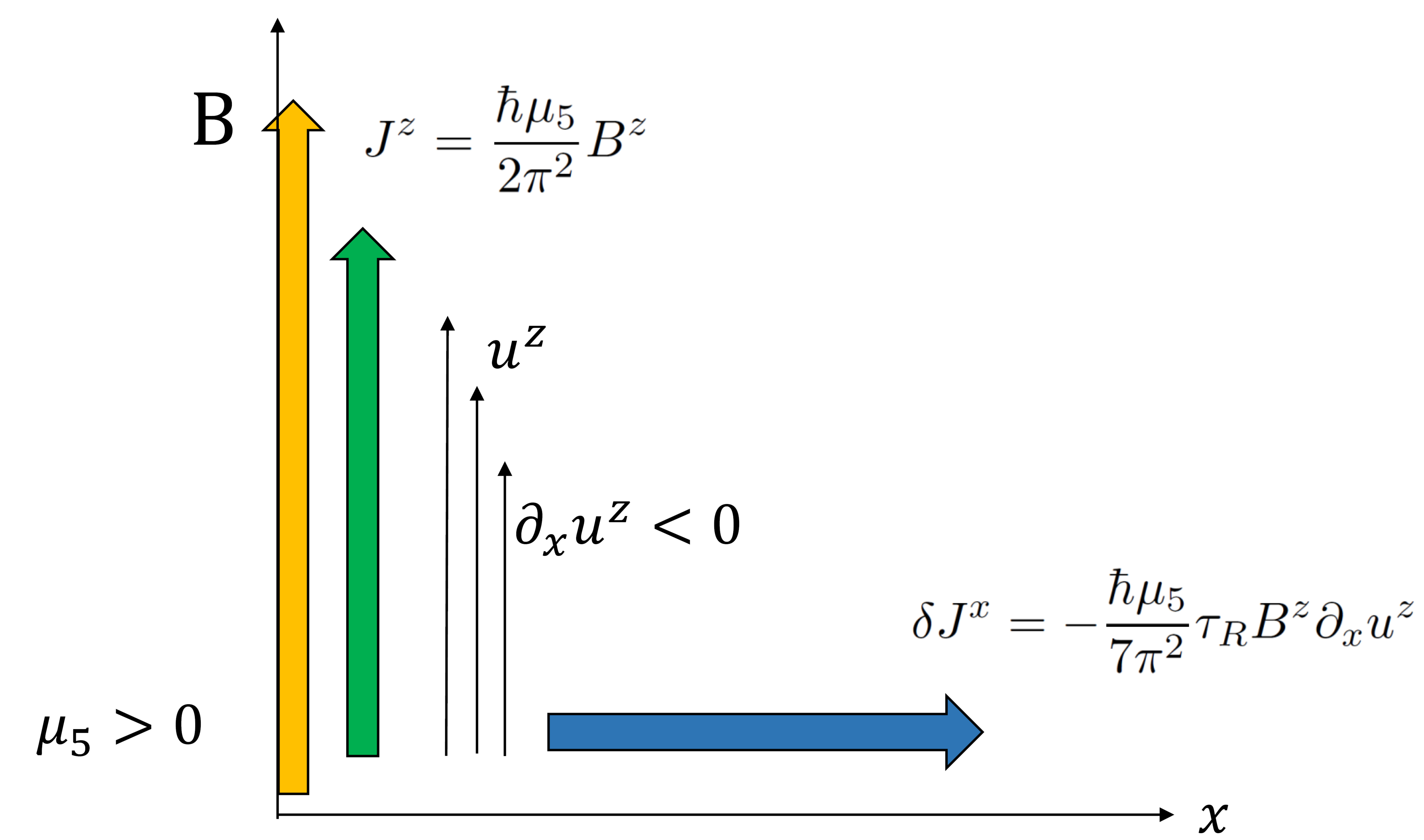}}
		\caption{A schematic figure for a CME Hall current led by shear corrections.}\label{HCME}
	\end{center}
\end{figure}

Finally, the non-equilibrium second-order anomalous corrections upon the charged current for right-handed fermions are summarized in Table 1, in which various anomalous Hall currents at $\mathcal{O}(\hbar\partial^2)$ triggered by electric fields and temperature/chemical-potential gradients are found \cite{Hidaka:2017auj}. Furthermore, the CME/CVE receive viscous corrections \cite{Hidaka:2018ekt}, where we define  $P^{\mu\nu}\equiv\eta^{\mu\nu}-u^{\mu}u^{\nu}$ and $\theta\equiv\partial\cdot u$ as the bulk strength and 
$\pi^{\mu\nu}\equiv P^{\mu}_{\rho}P^{\nu}_{\sigma}(\partial^{\rho}u^{\sigma}
+\partial^{\sigma}u^{\rho}-2\eta^{\rho\sigma}\theta/3)/2$ as the shear strength. The detials of transport coefficients can be found in Refs.\cite{Hidaka:2017auj,Hidaka:2018ekt}, while the non-equilibrium corrections are linear to $\tau_R$ and hence interaction-dependent. Also, such anomalous second-order transport is dissipative as opposed to the classical one. The viscous correction on CME in fact originates from a time-dependent magnetic field led by the shear/bulk strengths through the Bianchi identity, which is a reminiscence of the AC conductivity of CME and not surprised to depend on interactions \cite{Kharzeev:2016sut}. In Fig. \ref{HCME}, we further include the contributions from left-handed fermions and illustrate a schematic picture for the shear correction on CME at $\mu\ll T$ led by the fluid-velocity gradient, where $\mu_5=(\mu_R-\mu_L)/2$ is an axial chemical potential. Such a correction could engender a CME Hall current perpendicular to the magnetic field.         
\begin{table}
\begin{center}\label{tab1}
	\setlength{\tabcolsep}{6pt}
	\renewcommand{\arraystretch}{1.5}
	\begin{tabular}{ | l | l |  p{10cm} |}
		\hline
		$J^{\mu}_{Q\perp}$ & $\mathcal{O}(\partial)$ & $\mathcal{O}(\partial^2)$ \\ \hline
		$\mathcal{O}(\hbar)$ & $\sigma_BB^{\mu}+\sigma_{\omega}\omega^{\mu}$ 
		& $\tau_R\big[\epsilon^{\mu\nu\alpha\beta}u_{\nu}\big(\hat{\gamma}_E\partial_{\alpha} E_{\beta}+\hat{\gamma}_{\mu}E_{\alpha}\partial_{\beta} \mu+\hat{\gamma}_{T} E_{\alpha}\partial_{\beta} T+\hat{\gamma}_{T\mu}(\partial_{\alpha} T)(\partial_{\beta}\mu)\big)$
		$+\delta\hat{\sigma}_{BL}\theta B^{\mu}
			+\delta\hat{\sigma}_{BH}\pi^{\mu\nu} B_{\nu}$
		$+\delta\hat{\sigma}_{\omega L}\theta \omega^{\mu}
			+\delta\hat{\sigma}_{\omega H}\pi^{\mu\nu} \omega_{\nu}\big]$
		\\
		\hline
	\end{tabular}
	\caption{quantum corrections upon the charged current for right-handed fermions.}
\end{center}
\end{table}

\section{Outlook}
The WF-formalism and CKT can be further applied to study other transport properties of chiral fluids such as relativistic angular momenta and polarization \cite{Yang:2018lew}. In addition, for future phenomenological studies in HIC, we have to incorporate dynamical gluons and realistic collisions in QGP. On the other hand, it is also important to generalize the current formalism beyond the weak-field limit, which then entails further exploration in WF with higher-order quantum ($\hbar$) corrections.

Acknowledgments:
D. Y. was supported by the RIKEN Foreign Postdoctoral Researcher program.
Y. H. was partially supported by Japan Society of Promotion of Science (JSPS), Grants-in-Aid for Scientific Research
(KAKENHI) Grants No. 15H03652, 16K17716, and 17H06462, and also by RIKEN iTHES Project and iTHEMS Program. S.P. was supported by
JSPS post-doctoral fellowship for foreign researchers.
D. Y. was supported by the RIKEN Foreign Postdoctoral Researcher program.




\bibliographystyle{elsarticle-num}
\bibliography{DL_Yang_QM18.bbl}







\end{document}